\documentclass[12pt]{iopart}

\usepackage{iopams}

\usepackage{array}
\usepackage{graphicx}
\usepackage{braket}

\newcommand{\eqref}[1]{(\ref{#1})}

\begin{document}

% \title{The Interaction Quench in the 2D Hubbard model from a semiclassical perspective}
\title[Systematic large flavor fTWA approach to interaction quenches]{Systematic large flavor fTWA approach to interaction quenches in the Hubbard model}


\author{Alexander Osterkorn}
\address{Jožef Stefan Institute, Jamova 39, 1000 Ljubljana, Slovenia}
\address{Institute for Theoretical Physics,  Georg-August-Universit\"at G\"ottingen, Friedrich-Hund-Platz 1 - 37077  G\"ottingen, Germany}
\ead{alexander.osterkorn@ijs.si}
%\thanks{}

\author{Stefan Kehrein}
\address{Institute for Theoretical Physics,  Georg-August-Universit\"at G\"ottingen, Friedrich-Hund-Platz 1 - 37077  G\"ottingen, Germany}


\date{\today}

\begin{abstract}
We study the nonequilibrium dynamics after an interaction quench in the two-dimensional Hubbard model using the recently introduced fermionic truncated Wigner approximation (fTWA).
To assess the range of validity of the method in a systematic way,
we consider the SU($N$) Hubbard model with the fermion degeneracy $N$ as a natural semiclassical expansion parameter.
Using both a numerical and a perturbative analytical approach we show that fTWA is exact at least up to and including the prethermalization dynamics.
We discuss the limitations of the method beyond this regime.
\end{abstract}
%\submitto{\jpa}

%\keywords{}

\maketitle

\section{Introduction\label{sec:introduction}}

The dynamics of quantum systems out-of-equilibrium~\cite{Polkovnikov2011} is a very active field of research that offers a lot of open fundamental questions as well as many perspectives for technological applications.
The research is strongly driven by better and better possibilities to realize quantum mechanical model systems with ultracold gas experiments~\cite{Gross2017} and by the advancement of time-resolved experimental techniques in solid state physics~\cite{Dombi2020}.
In the latter context layered two-dimensional strongly correlated materials like the transition metal dichalcogenides~\cite{Manzeli2017} are currently moving into the center of interest.
In time-resolved angle-resolved photoemission spectroscopy (trARPES), one of the main experimental techniques to unravel the microscopic structure of such materials, the response of the electronic system to the application of a strong laser pulse is measured.
This in turn requires reliable theoretical simulations of such setups in order to link the experimental observations with microscopic models \cite{Freericks2009a,Freericks2015}.
However, theoretical simulations of the light-induced quantum dynamics in correlated systems are very challenging due to the lack of a numerical or analytical method that is valid both for a broad range of systems and over long periods of time~\cite{Eckstein2009}.
Established approaches include tensor-network based methods~\cite{Paeckel2019}, the non-equilibrium extension of dynamical mean-field theory (DMFT)~\cite{Freericks2006,Aoki2014} as well as perturbative schemes.
While the first are very powerful for one-dimensional quantum systems, their usefulness is restricted for time-dependent problems in 2d.
Nonequilibrium DMFT is believed to work well for three-dimensional materials,
its reliability in only two spatial dimensions is not clear because of the approximation of the lattice as high dimensional and the lack of systematic error bounds.
Perturbative approaches are applicable to many systems but are limited to weak interactions, can suffer from secular terms~\cite{Hackl2008} or may not be able to treat explicitly time-dependent Hamiltonians.

\begin{figure}
	\centering
	\includegraphics[width=\textwidth]{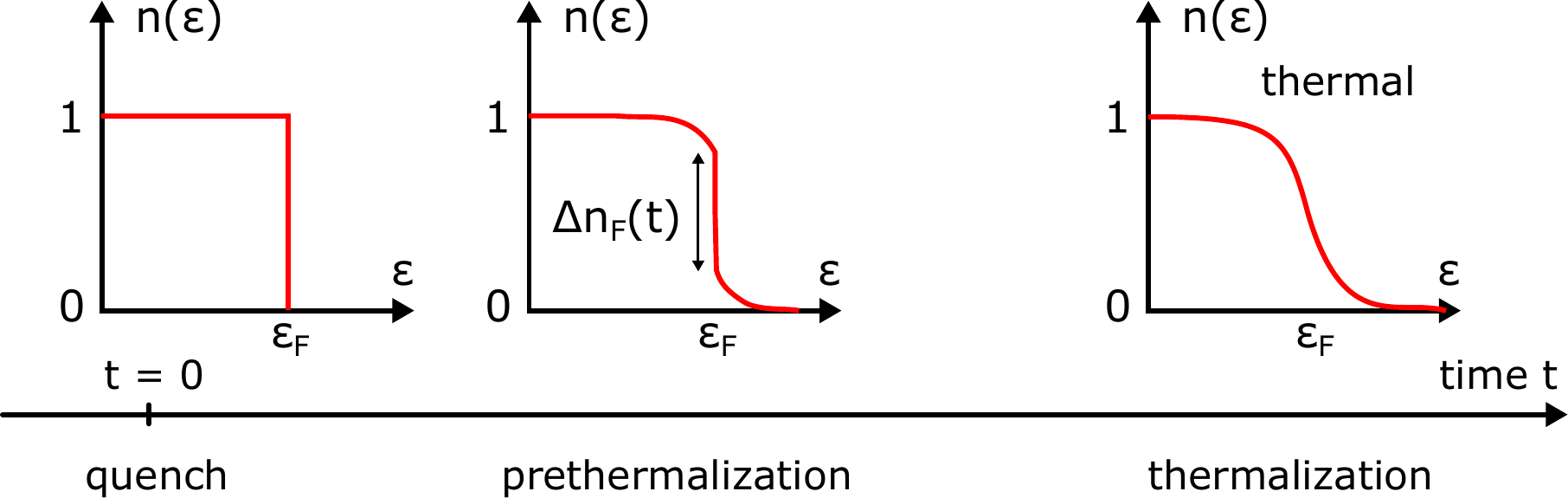}%
	\caption{Sketch of the basic phenomenology of the interaction quench in the Hubbard model for quenches to weak interaction strengths. The diagrams show schematically the electronic occupation numbers $n(\epsilon_k)$ as they evolve with time. Starting from the Fermi-Dirac distribution at zero temperature, dephasing dynamics leads to a reduction of the Fermi surface discontinuity.
	The so-obtained \emph{prethermal} distribution function remains approximately constant on an intermediate timescale before thermalization becomes efficient and the occupation numbers ultimately evolve to a smooth thermal distribution.
	\label{fig:hub_quench-sketch}}
\end{figure}

In theoretical quantum optics, semiclassical descriptions have shown to be useful to simulate the dynamics of interacting bosons~\cite{Sinatra2002,Polkovnikov2002,Polkovnikov2003}.
Unfortunately, much less experience with semiclassics for fermions exists and only recently some method development in this direction was reported~\cite{Ayik2008,Lacroix2014,Lacroix2014a,Davidson2017}.
These developments naturally raise the question which quantum effects are captured by a semiclassical treatment of lattice fermions and if such an approach is useful in two (and higher) spatial dimensions.
In this text we adopt one of these recent developments, the fermionic truncated Wigner approximation (fTWA) and apply it to the well-understood problem of the quench from zero to weak interaction strength in the Hubbard model~\cite{Moeckel2008,Moeckel2009,Moeckel2010,Eckstein2009,Hamerla2014}, which we implement on a square lattice.
The interaction quench problem is very suitable for method benchmarking since it shows correlation-induced physics on well-separated timescales.
Fig.~\ref{fig:hub_quench-sketch} shows a sketch of the basic phenomenology:
At initial time, the occupation numbers $n(\epsilon_k)$ of the electrons follow the box-shaped Fermi-Dirac distribution function for zero temperature.
After the sudden switch-on of a weak interaction, the electrons become dressed and the early-time dynamics is characterized by dephasing into the quasi-particle basis.
During the dephasing dynamics, the discontinuity $\Delta n_\textrm{\scriptsize F} = n(\epsilon_{k_\textrm{\scriptsize F}} - 0) - n(\epsilon_{k_\textrm{\scriptsize F}} + 0)$ at the Fermi surface shrinks but remains nonzero.
The timescale of the dephasing dynamics scales like $\sim U^{-2}$,
while the scattering of the quasi-particles, which ultimately leads to thermalization, happens on a slower timescale $\sim U^{-4}$.
This timescale separation implies the formation of a characteristic “prethermalization plateau” in the time dependent $\Delta n_\textrm{\scriptsize F}(t)$ before the thermalization dynamics dominates.

The guiding question of this paper is therefore, which regimes of the above-described dynamics can be captured by the semiclassical approach.
After introducing the method, we will combine it with an explicit semiclassical expansion parameter and present perturbative analytical as well as numerical results in order to shed some light on the range of validity of the fTWA method.

\section{Semiclassical quantum dynamics}

\subsection{General framework}

The concept of semiclassical dynamics encompasses a number of approaches that replace the full quantum mechanical description of a physical system by a classical description and allow to incorporate quantum effects in a controlled way.
A typical way to construct such theories is a formal expansion of the quantum theory in $\hbar$.
The leading order contribution as $\hbar \rightarrow 0$ yields a description in terms of classical variables.
In particular, quantum Hamiltonians are converted to classical Hamiltonian systems.
An intuitive understanding of this stems from the trivialization of commutator relations like $[\hat q, \hat p] = i \hbar$ in this limit.
Many quantum systems contain a natural expansion parameter that can be used to define an "effective $\hbar$".
Among the most prominent examples are large-spin and large-$N$ expansions~\cite{Yaffe1982,Bickers1987} as well as expansions in the mode occupation of Bose-Einstein condensates~\cite{Gardiner1997}.
The resulting classical theory is often interpretable as a mean-field description of the original quantum theory.
In the case of interacting bosons, for instance, the leading order classical description is given by the Gross-Pitaevskii equation, which is as well obtained from a mean-field decoupling of the interaction term.

Arguably the most prominent approach to add quantum corrections in a systematic manner is the truncated Wigner approximation (TWA).
Working in the phase space formulation of quantum mechanics, it can be obtained from a systematic expansion of the von Neumann equation in $\hbar$ and a subsequent truncation to order $\hbar$ \cite{Polkovnikov2010}.
Alternatively, a derivation from the path integral representation of the Keldysh formalism is possible \cite{Polkovnikov2003}.
The idea at the heart of TWA is that of an effective Liouville dynamics.
Using a set $\underline{\rho}$ of phase space variables that fully characterize the physical system, like coordinate and momentum, spin components or bosonic modes, states are described in terms of their Wigner quasi-probability distributions $W(\underline{\rho})$.
Their time-evolution, in turn, is governed by the flow generated from the Hamiltonian $H_\textrm{\scriptsize class}$ that corresponds to the zeroth order in $\hbar$ classical description
\begin{equation}
 i \partial_t W(\underline{\rho}) \simeq \left\{ H_\textrm{\scriptsize class}(\underline{\rho}), W(\underline{\rho}) \right\} .
\end{equation}
This effective Liouville equation gives rise to a prescription for the evaluation of operator expectation values via the
statistical averaging over trajectories in phase space
\begin{equation}
\big\langle \hat O(t) \big\rangle \simeq \int \textrm{d}\underline{\rho} \; W(\underline{\rho}) O_\textrm{\scriptsize W}(\underline{\rho}(t)) ,
\end{equation}
Here, $\underline{\rho}(t)$ is time-evolved according to the Hamiltonian equations of motion for $H_\textrm{\scriptsize class}$ and $O_\textrm{\scriptsize W}$ denotes the classical analogue of the quantum mechanical operator $\hat O$, i.e. its Weyl symbol \cite{Polkovnikov2010}.

\subsection{Semiclassics for fermions}

While the TWA method as described above was successfully applied to bosonic systems, it was only recently extended to fermionic degrees of freedom \cite{Davidson2017}.
The extension is called fermionic TWA (fTWA) and defines a set of phase space coordinates by making use of the so($2n$) commutator structure of the fermionic bilinears $\hat \rho_{\alpha\beta} = c_\alpha^\dagger c_\beta - \frac{1}{2} \delta_{\alpha\beta}$ and $\hat \tau_{\alpha\beta} = c_\alpha c_\beta$.
fTWA was used to study the thermalization and echo dynamics in  SYK models~\cite{Davidson2017,Schmitt2019} as well as the non-equilibrium dynamics in disordered models \cite{Sajna2020,Iwanek2023,Kaczmarek2023}.
An equivalent method was proposed earlier in a different context under the name “stochastic mean-field approach”~\cite{Ayik2008,Lacroix2014,Lacroix2014a}.

Within fTWA, the operators $\hat \rho_{\alpha\beta}$ and $\hat \tau_{\alpha\beta}$ are replaced by their associated classical phase space variables $\rho_{\alpha\beta}$ and $\tau_{\alpha\beta}$, i.~e. their Weyl symbol in the context of the phase space formulation of quantum mechanics.
The semiclassical time-evolution equations are derived using a mean-field decoupling of the interaction term in the fermionic many-body Hamiltonian.
For the application to the Hubbard model in this text only the operators $\hat \rho_{\alpha\beta}$ with an index set $\alpha = i\sigma$ need to be considered, where $i$ denotes the lattice site $i$ and $\sigma$ is a spin index.
The Wigner function is typically constructed as a probability distribution function with means and connected covariances determined from the respective expectation values of the quantum initial state:
\begin{equation}\begin{array}{*2{>{\displaystyle}l}}
 \langle \rho_{i\sigma, j\sigma} \rangle_W & \mathop{=}\limits^{!}\quad \langle \hat \rho_{i\sigma, j\sigma} \rangle_\textrm{\scriptsize QM}, \\
 \left\langle \rho_{i\sigma, j\sigma}^\ast \rho_{k\tau, l\tau} \right\rangle_W^\textrm{c.c.}
 &\mathop{=}\limits^{!}\quad \frac{1}{2} \left\langle \big\{ \hat \rho_{i\sigma, j\sigma}^\dagger, \hat \rho_{k\tau, l\tau} \big\} \right\rangle_\textrm{\scriptsize QM}^\textrm{c.c.} ,
 \label{eq:ftwa_gauss_wigner_params}
\end{array}\end{equation}
where $\langle a b \rangle^\textrm{c.c.} = \langle a b \rangle - \langle a \rangle \langle b \rangle$ denotes connected correlations.
The simplest choice for $W$ is a Gaussian distribution,
although other choices like the two-point function $P_\textrm{\scriptsize TP}(x) = \frac{1}{2} \big( \delta(x - \mu - \sigma) + \delta(x - \mu + \sigma) \big)$ have shown to be advantageous for some applications~\cite{Ulgen2019}.

\subsection{Large-$N$ as a semiclassical limit for lattice fermions}

Despite the fact that fTWA often yields good agreement with exact calculations on short and intermediate time scales \cite{Davidson2017} it is essentially an uncontrolled approximation.
This is a consequence of the lack of a natural semiclassical expansion parameter for fermions, since -- in contrast to bosons -- occupation numbers are bounded.
One possibility to systematically improve the validity of the method is to tune the range of the fermionic interactions \cite{Sajna2020} from short-range up to very long-range.

In this text, we combine fTWA with a SU($N$)-symmetric formulation of the Hubbard model that keeps the short-rangedness of the interaction but instead increases the dimension $N$ of the local electronic state space.
Such approaches are common in equilibrium statistical physics, e.~g. for frustrated magnets \cite{Read1991,Sachdev1991}, intermediate valence systems \cite{Newns1987} and correlated lattice electrons \cite{Affleck1988,Osterkorn2022}.
In addition, the application of large-$N$ techniques for  non-equilibrium physics is becoming more popular \cite{Kronenwett2011,Weidinger2017,Walz2018}.
Furthermore, the experimental realization of models with values of $N$ up to 10 is possible in an ultracold atom setting \cite{Gorshkov2010,Choudhury2020}, which provides an additional motivation for the approach.

In this paper, we consider fermionic operators $c_{i\alpha}^\dagger, c_{i\alpha}$ with $\alpha = 1, \dots, N$ different spin states (flavors).
Within fTWA we may now define a set of flavor-averaged phase space variables
\begin{equation}
 \rho_{ij} = \frac{1}{N} \sum_{\alpha = 1}^{N} \rho_{i\alpha, j\alpha} .
 \label{eq:rho_flavor_avg}
\end{equation}
The commutation relations of the corresponding quantum mechanical operators collect an additional factor of $1/N$:
\begin{equation}
 [ \hat \rho_{ij}, \hat \rho_{mn} ] = \frac{1}{N} ( \delta_{jm} \hat \rho_{in} - \delta_{in} \hat \rho_{mj} )
 \label{eq:su_n_ftwa_commrel}
\end{equation}
This illustrates the semiclassical nature of the parameter $N$.
In the limit $N \rightarrow \infty$ the commutation relations  \eqref{eq:su_n_ftwa_commrel} are trivialized and the operators effectively behave like classical variables.

\section{Model and method setup}

In the following, we study the time evolution of the square lattice Fermi sea, which is the ground state of the non-interacting model, under a Hubbard Hamiltonian with $U > 0$:
\begin{equation}
H_0 = \sum_k \epsilon_k \hat c_k^\dagger \hat c_k ~\Rightarrow~ H = \sum_k \epsilon_k \hat c_k^\dagger \hat c_k + U \sum_i \hat n_{i\downarrow} \hat n_{i\uparrow} ,
\end{equation}
where $\epsilon_k = -2t_\textrm{\scriptsize h} \big( \cos(k_x) + \cos(k_y) \big)$.
%This quench protocol has intensively been studied both analytically \cite{Moeckel2008,Moeckel2009,Moeckel2010} and numerically \cite{Eckstein2009,Hamerla2014}.
%Among its hallmark features are the existence of a prethermalizion regime at time scales before the thermalization dynamics dominates.
The SU($N$)-invariant version of the Hubbard model reads as follows \cite{Marston1989}:
\begin{equation}
 \hat H = - t_\textrm{\scriptsize h} \sum_{\langle i, j \rangle} \sum_\alpha c_{i\alpha}^\dagger c_{j\alpha} + \textrm{H.c.} + \frac{U}{N} \sum_i \left( \sum_\alpha c_{i\alpha}^\dagger c_{i\alpha} - \frac{N}{2} \right)^2
\end{equation}
The structure of the Hamiltonian allows for a natural representation in terms of the flavor-averaged $\hat \rho$-operators~\eqref{eq:rho_flavor_avg}:
\begin{equation}\begin{array}{*2{>{\displaystyle}l}}
 \hat H = N \bigg[ - t_\textrm{\scriptsize h} \sum_{\langle ij \rangle} \left( \hat \rho_{ij} + \hat \rho_{ji} \right) + U \sum_i \hat \rho_{ii}^2 \bigg]
 \label{eq:su_n_ftwa_hamiltonian}
\end{array}\end{equation}
In addition, the use of such flavor-averaged phase space variables resolves an ambiguity in the classical representation of the interaction term which is due to the quantum mechanical identity $n_{i\alpha}^2 = n_{i\alpha}$ for fermions.
For $N = 2$, the semiclassical Hamiltonian
\begin{equation}
 H_\textrm{\scriptsize int}^{(I)} = U \sum_i \left( \rho_{i\uparrow, i\uparrow} + \frac{1}{2} \right) \left( \rho_{i\downarrow, i\downarrow} + \frac{1}{2} \right)
 \label{eq:ftwa_ham_decoupl_naive}
\end{equation}
would be quantum mechanically, but is not semiclassically equivalent to the representation derived from the SU($N$)-invariant Hamiltonian
\begin{equation}
 H_\textrm{\scriptsize int}^{(II)} = \frac{U}{2} \sum_i \left( \rho_{i\uparrow,i\uparrow} + \rho_{i\downarrow, i\downarrow} \right)^2 .
 \label{eq:ftwa_ham_decoupl_su_n}
\end{equation}
However, for the problem considered in this text, we did not observe differing results between the two representations.
In other contexts~\cite{Davidson2017,Sajna2020}, a specific choice of the representation has turned out to yield better numerical results than other choices.

The equations of motion for the phase space variables $\rho_{ij}$ can be obtained from the classical Hamiltonian formalism \cite{Davidson2017} upon mean-field decoupling $\hat \rho_{ii}^2 \rightarrow \rho_{ii}^2$.
Equivalently, they follow from the Heisenberg equations of motion corresponding to \eqref{eq:su_n_ftwa_hamiltonian} in the limit $N \rightarrow \infty$,
\begin{equation}
 i \partial_t \rho_{ij} = - t_\textrm{\scriptsize h} \sum_{a(j)} \rho_{i, a(j)}
 + t_\textrm{\scriptsize h} \sum_{a(i)} \rho_{a(i), j}
 + 2 U (\rho_{jj} - \rho_{ii}) \rho_{ij} .
\end{equation}

The equilibrium ground state of the model with $U = 0$ is given by the $N$-flavor Fermi sea $\ket{\textrm{FS}} = \prod_{\alpha, |\vec k| \leq k_F} c_{\vec k \alpha}^\dagger \ket{0}$ whose initial data \eqref{eq:ftwa_gauss_wigner_params} in momentum space we can readily calculate:
\begin{equation}\begin{array}{*2{>{\displaystyle}l}}
\big\langle \hat \rho_{k l} \big\rangle = \big\langle \hat \rho_{k l}^\dagger \big\rangle &= \delta_{k, l} \left( n_k - \frac{1}{2} \right), \\
 \frac{1}{2} \Big\langle \big\{ \hat \rho_{k l},  \hat \rho_{s p} \big\} \Big\rangle^\textrm{c.c.} &= \frac{1}{2N} \delta_{k p} \delta_{l s} \left( n_k + n_l - 2 n_k n_l \right) .
\end{array}\end{equation}
As $N \rightarrow \infty$, the Hubbard interaction $U$ in \eqref{eq:su_n_ftwa_hamiltonian} merely plays the role of a shift of the
chemical potential such that non-trivial dynamics after the interaction quench can only occur at finite $N$.

% semiclassical equation of motion:
% \begin{align} \begin{split}
%  i \partial_t \rho_{i\sigma, j\sigma} = &-t \sum_{a(j)} \rho_{i\sigma, a(j)\sigma} + t \sum_{a(i)} \rho_{a(i)\sigma, j\sigma} \\
%  &+ U (\rho_{j\bar\sigma, j\bar\sigma} - \rho_{i\bar\sigma, i\bar\sigma}) \rho_{i\sigma, j\tau}
%  \label{eq:ftwa_spin_eom_pos}
% \end{split} \end{align}

% in momentum space:
% \begin{align} \begin{split}
%  &i \partial_t \rho_{k \sigma, l \sigma} = - \left( \epsilon(\vec k) - \epsilon(\vec l) \right) \rho_{\vec k \sigma, \vec l \sigma} \\
%  &+ \frac{U}{V} \sum_{\vec s, \vec p} \left[ \rho_{(\vec p + \vec s - \vec l) \bar\sigma, \vec p \bar\sigma} \rho_{\vec k \sigma, \vec s \sigma} - \rho_{(\vec p + \vec k - \vec s) \bar\sigma, \vec p \bar\sigma} \rho_{\vec s \sigma, \vec l \sigma} \right]
%  \label{eq:ftwa_spin_eom_mom}
% \end{split} \end{align}

\section{Results for the SU($N$) fTWA}

\subsection{Perturbative treatment of the e.o.m.}

For weak Hubbard interaction strengths $U \ll t_\textrm{\scriptsize h}$ one can treat the classical equations of motion perturbatively and evaluate all expectation values with respect to the Gaussian Wigner function by hand.
In order to do so, it is advantageous to work with the equations in momentum space.
Using the Fourier transform
\begin{equation}
 \rho_{ij} = \sum_{kl} \textrm{e}^{i (k r_i - l r_j)} \rho_{kl} ,
\end{equation}
one obtains the equations of motion in momentum space,
\begin{equation}\begin{array}{*2{>{\displaystyle}l}}
 i \partial_t \rho_{kl} = &- \left( \epsilon_k - \epsilon_l \right) \rho_{kl} \\
 &+ \frac{2U}{V} \sum_{sp} \left( \rho_{p+s-l,p} \rho_{ks} - \rho_{p+k-s,p} \rho_{sl} \right) .
\end{array}\end{equation}

A naive perturbative expansion of these equations in $U$ is only valid up to times $\mathcal{O}(t^2)$.
In order to avoid restricting secular terms we switch to an interaction picture representation of the equations of motion by incorporating the free time-evolution into the variables $\tilde{\rho}_{kl} = \textrm{e}^{-i (\epsilon_k - \epsilon_l) t} \rho_{kl}$ where
$\Delta\epsilon_{pks} = \epsilon_{p+k-s} + \epsilon_s - \epsilon_p - \epsilon_k$.
This yields
\begin{equation}
 i \partial_t \tilde{\rho}_{kl} = \frac{2U}{V} \sum_{sp} \bigg[ \textrm{e}^{i \Delta\epsilon_{psl} t} \tilde{\rho}_{p+s-l,p} \tilde{\rho}_{ks}
 - \textrm{e}^{i \Delta\epsilon_{pks} t} \tilde{\rho}_{p+k-s,p} \tilde{\rho}_{sl} \bigg] .
 \label{eq:su_n_ftwa_eom_mom_int}
\end{equation}

We may now expand the variables order by order in $U$
\begin{equation}
 \tilde \rho_{k l} = \tilde \rho^{(0)}_{k l} + \tilde \rho^{(1)}_{k l} \cdot U + \tilde \rho^{(2)}_{k l} \cdot U^2 + \dots .
\end{equation}
Inserting the ansatz into \eqref{eq:su_n_ftwa_eom_mom_int} yields a hierarchy of equations with increasing orders of $U$.
The zeroth order contribution is constant in time, $i \partial_t \tilde{\rho}^{(0)}_{k l} = 0$.
This fact allows to explicitely integrate all time dependencies in the equation for $\tilde{\rho}^{(1)}_{k l}$.
In a last step, all expectation values of products of $\tilde{\rho}^{(0)}_{k l}$ are evaluated using the Gaussian Wigner function.
Successive application of this scheme results in an iterative procedure to solve for the dynamics to all orders of $U$.
The elastic contributions with $\Delta\epsilon_{pp'k} = 0$ lead to diverging energy denominators and cannot be treated in this perturbative approach as they would produce secular terms.
In the long-time limit we expect that these terms give rise to dynamics governed by a quantum Boltzmann equation~\cite{Moeckel2008}. 
More details of the calculation are shown in~\ref{app:perturb}.
We solved the hierarchy up to order $U^2$ and obtained the following results:
%\begin{equation}\begin{array}{*2{>{\displaystyle}l}}
% \tilde \rho^{(1)}_{k l}(t) &= 0 \\
% \tilde \rho^{(2)}_{k l}(t) &= - \delta_{k, l} \frac{16}{N V^2} \sum_{\mathclap{\substack{pp' \\ \Delta\epsilon_{pp'k} \neq 0}}} \frac{\sin^2 \left( \frac{\Delta\epsilon_{pp'k}}{2} t \right)}{(\Delta\epsilon_{pp'k})^2} J_{pp'k}
% \label{eq:ftwa_perturb_res}
%\end{array}\end{equation}
\begin{equation}\begin{array}{*2{>{\displaystyle}l}}
	\tilde \rho^{(1)}_{k l}(t) &= 0 \\
	\tilde \rho^{(2)}_{k l}(t) &= - \delta_{k, l} \frac{16}{N V^2} \sum\limits_{pp', \Delta\epsilon_{pp'k} \neq 0} \frac{\sin^2 \left( \frac{\Delta\epsilon_{pp'k}}{2} t \right)}{(\Delta\epsilon_{pp'k})^2} J_{pp'k} ,
	\label{eq:ftwa_perturb_res}
\end{array}\end{equation}
where
\begin{equation}
 J_{p p' k} = n_k n_{p+p'-k} (1-n_p) (1-n_{p'}) - n_p n_{p'} (1 - n_k) (1 - n_{p + p' -k})
\end{equation}
is a phase space factor.
These results agree precisely with those obtained from unitary perturbation theory \cite{Moeckel2008} for the prethermal dynamics.
It is worth noting that via the sampling of the initial conditions the truncated Wigner approach with time-local mean-field equations of motion is able to reproduce the correlation-induced prethermalization dynamics.
Physically, this dynamics at the perturbative order $U^2$ describes electronic dephasing, while the dynamics beyond this regime is due to the scattering of quasiparticles.

\subsection{Numerical results}

In order to study the quench dynamics numerically, we implemented Hamilton's equations of motion using the odeint library \cite{Ahnert2011} and the Armadillo library \cite{Sanderson2016,Sanderson2018}.
To avoid the accumulation of numerical errors,
Welford's algorithm is used for checkpointing \cite{Schubert2018}.
Unless stated otherwise, we use a Gaussian Wigner function model.
To monitor the convergence of the simulation,
we made use of the fact that, due to the lattice symmetry, there are usually several momentum vectors $k$ that yield the same single-particle energy $\epsilon_k$.
Averaged over an infinite number of trajectories, observables like occupation numbers $n(\epsilon_{k_i}) = \rho_{k_i,k_i} + \frac{1}{2}$ should become identical at all momenta $k_i$ in such a set.
Therefore, we use the standard deviation of observables within these sets of $k$-values with equal band energies $\epsilon_k$ as a measure of convergence. 
We stopped sampling from the Wigner function when upon increasing the number of trajectoris the deviations between the results of observables (i.e. $\langle \rho_{kk} \rangle(t)$) at energy-identical $k$-values became small.
Especially for late times, the convergence with the number of trajectories can be very slow.
For values $U \ll t_\textrm{\scriptsize h}$, the numerical magnitude of expectation values is similar to the statistical noise such that the relative statistical error from the sampling is larger than for $U \sim t_\textrm{\scriptsize h}$.
In the former case we typically averaged over about $2 \cdot 10^5$ trajectories and in the latter case about $2 \cdot 10^4$ trajectories.

Two characteristic observables for the interaction quench dynamics are the jump $\Delta n(\epsilon_{k_\textrm{\scriptsize F}})$ of the momentum distribution $n_k = n(\epsilon_k)$ at the Fermi energy and the interaction energy $E_\textrm{\scriptsize int} \sim \langle n_i^2 \rangle$.
The first is directly related to the quasiparticle weight $Z$ \cite{Moeckel2008} and is equal to one for the initial Fermi-Dirac distribution with zero temperature.
The interaction energy is, in contrast to the mode-dependent $n_k$, a local quantity that is expected to relax during prethermalization to the equilibrium value of the post-quench Hamiltonian at the final temperature (determined by the amount of quench energy).
It provides a generalization of the double occupancy $d(t) = \langle n_{i\downarrow} n_{i\uparrow} \rangle$ for $N = 2$.
The conservation of total energy allows to compute the perturbative result for the change of the interaction energy at order $U^2$ from~\eqref{eq:ftwa_perturb_res}.

\begin{figure}
\centering
\includegraphics[width=\textwidth]{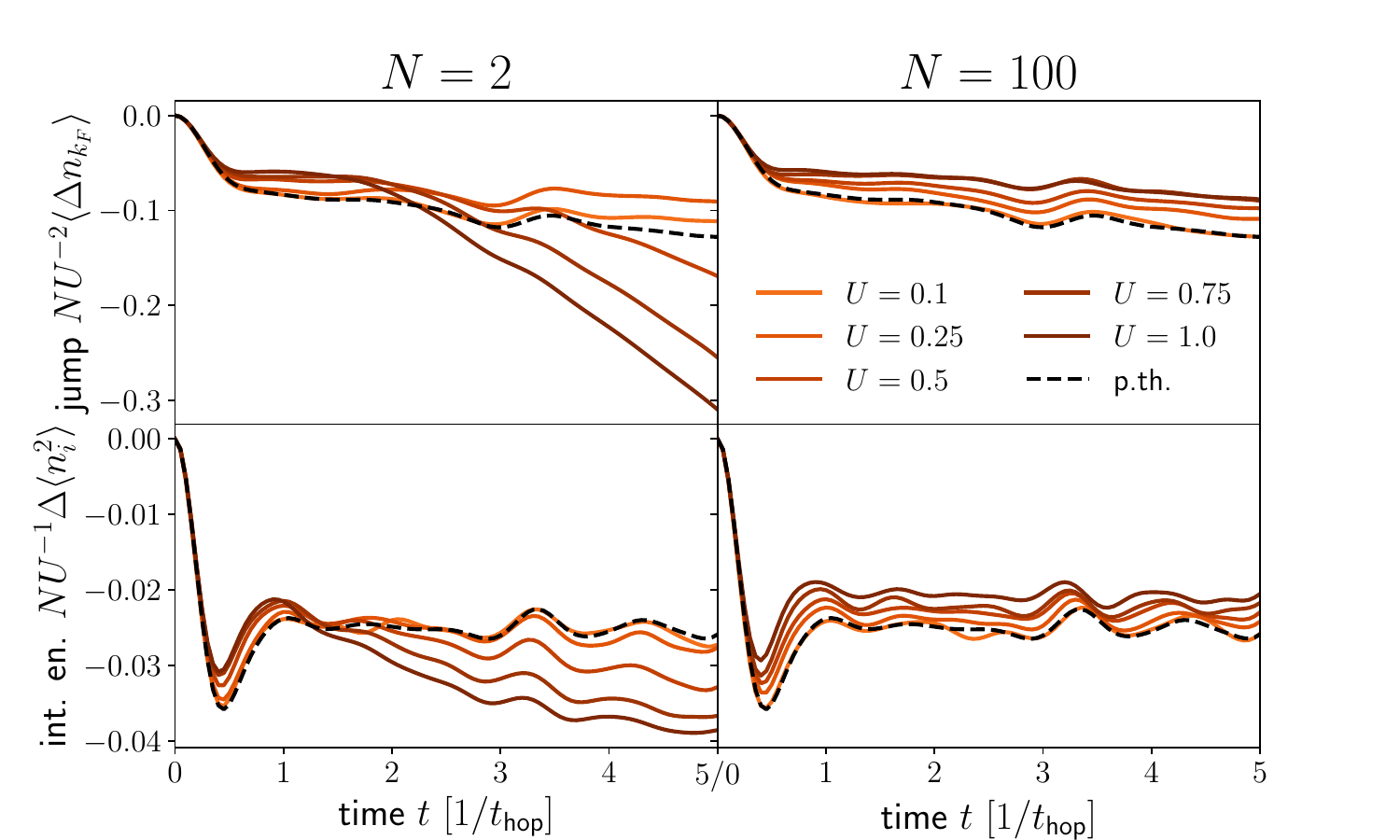}%
\caption{Semiclassical fTWA dynamics in a $10 \times 10$ square lattice Hubbard model at quarter filling ($N = 25$ particles) after a  quench to weak and intermediate values of $U$. The two columns display results at $N = 2$ and $N = 100$.
The top two panels present the jump in the momentum distribution at the Fermi energy, rescaled by a factor of $N/U^2$ that allows for a comparison to the result obtained from perturbation theory.
In the bottom row, the rescaled change of the interaction energy is shown. The dashed black curves are calculated using the perturbative result for the occupation numbers in combination with the conservation of the total energy throughout the dynamics. \label{fig:fermi_inten_N2_N100_L10}}
\end{figure}

\begin{figure}
\centering
\includegraphics[width=0.8\textwidth]{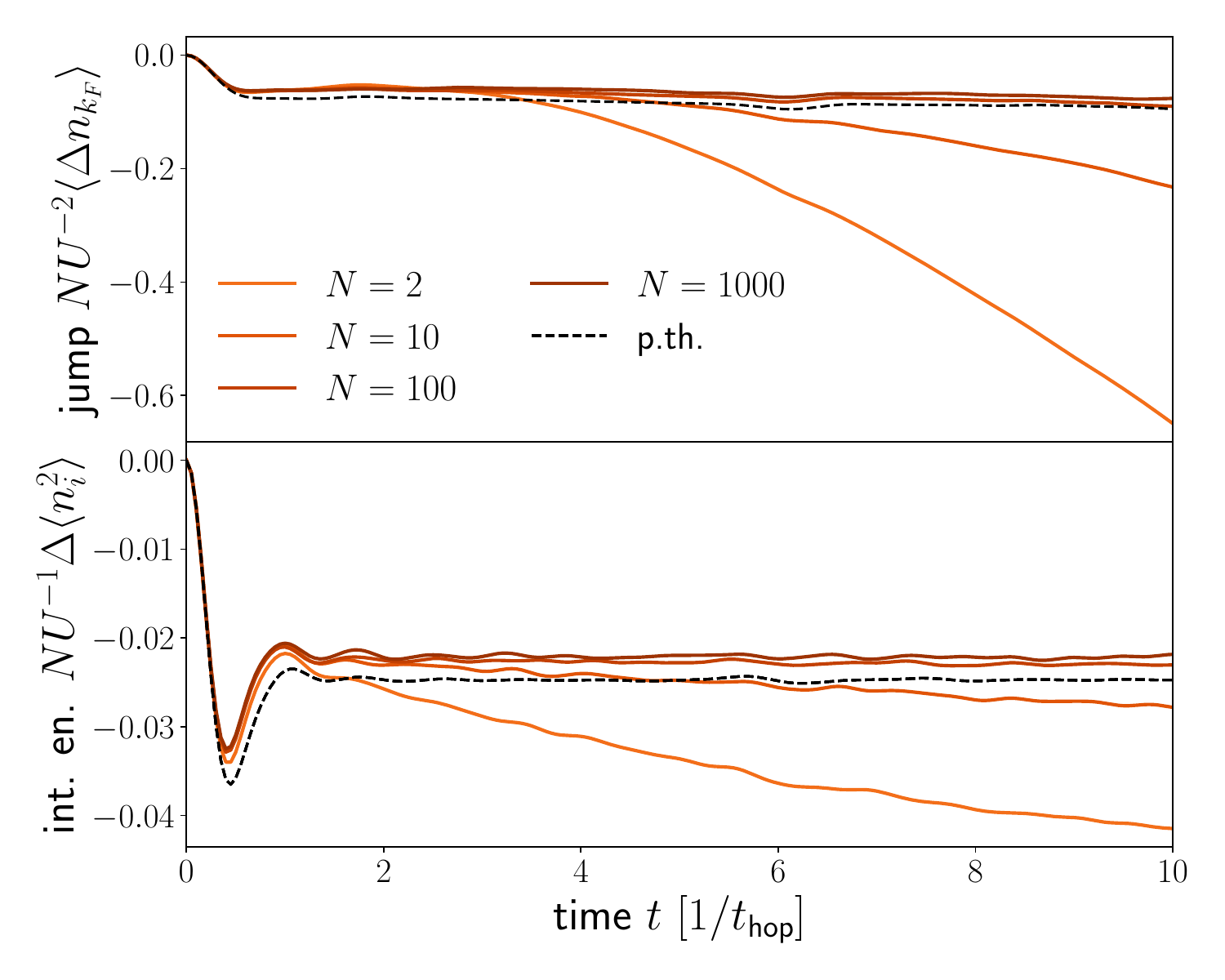}%
\caption{Fermi surface discontinuity of the occupation number $n(\epsilon_k)$ and interaction energy change $\sim \big( \langle n_i^2 \rangle(t) - \langle n_i^2 \rangle(0) \big)$ for a $20 \times 20$ square lattice Hubbard model at quarter filling ($N = 101$ particles) after an interaction quench to $U = 0.5 t_\textrm{\scriptsize h}$ for varying values of the fermion degeneracy $N$.\label{fig:fermi_inten_U5e-1_L20}}
\end{figure}

Since
% it is possible that 
prethermalization effects are suppressed at half filling in the thermodynamic limit in 2D~\cite{Hamerla2014}, we consider quarter filling in the following.
Fig.~\ref{fig:fermi_inten_N2_N100_L10} shows our numerical results for $\Delta n_{k_F}(t)$ and for the change of $\langle n_i^2 \rangle$ in a $10 \times 10$ system at two fixed values of the degeneracy parameter $N$,
whereas Fig.~\ref{fig:fermi_inten_U5e-1_L20} shows data for a $20 \times 20$ system at a fixed intermediate value of $U = 0.5 t_\textrm{\scriptsize h}$ and for varying $N$.
If $N = 2$, the SU($N$) model reduces to the conventional Hubbard model (with $H_\textrm{\scriptsize int}^{(II)}$).
We scale the obversables in the figures in a way that their order $U^2/N$ dynamics according to \eqref{eq:ftwa_perturb_res} would coincide.
This allows to better focus on the deviations from the perturbative result.
For weak $U$,
%in Fig. \ref{fig:fermi_inten_N2_N100_L10},
the dynamics at the Fermi edge agrees very well with the perturbative calculation.
For $U \sim t_\textrm{\scriptsize h}$ one can clearly expect a deviation from the perturbative result but it is noteworthy that, still, the overall shape of the curve does not change much for all the interaction strength values considered here.
This is in agreement with other exact numerical treatments of the interaction quench problem \cite{Eckstein2009}.
After the initial correlation build-up, a plateau is forming until, for $N = 2$, a further reduction of the discontinuity sets in. One can observe, in particular, that the width of the plateau is smaller for greater values of $U$, whereas for $N = 100$ the prethermalization plateau extends over a much longer times.
The onset of this reduction for varying $N$ can well be seen in Fig. \ref{fig:fermi_inten_U5e-1_L20}.
These results confirm the analytical calculation and show that the prethermal regime is indeed captured by fTWA.

\begin{figure}
\centering
\includegraphics[width=0.8\textwidth]{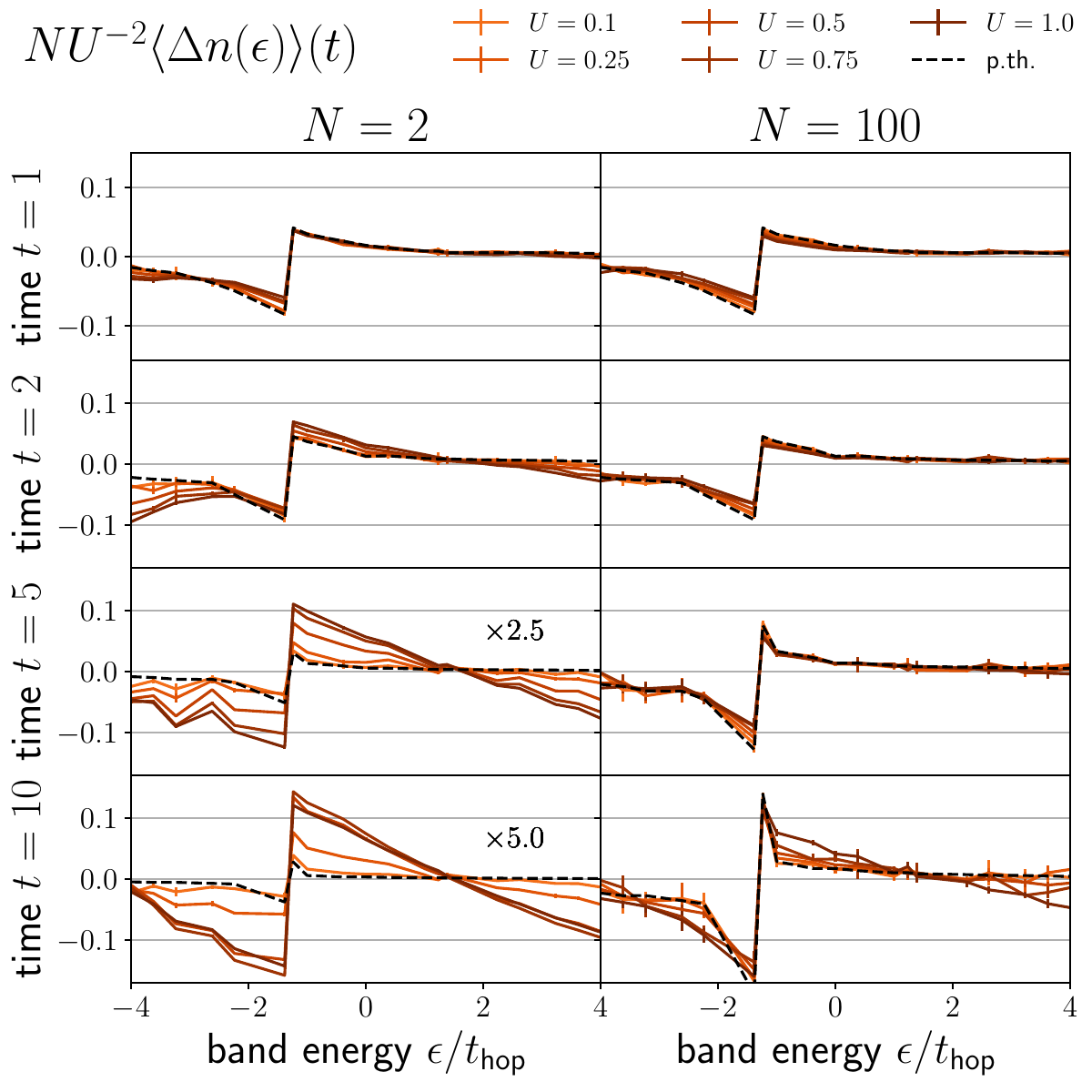}%
\caption{Electronic occupation numbers as a function of the band energy $\epsilon_k$ for a $10 \times 10$ square lattice Hubbard model at quarter filling ($N = 25$ particles) after a quench to interaction strengths $U \leq t_\textrm{\scriptsize h}$.
The two columns correspond to values of the fermion degeneracy $N$ of $2$ and $100$, respectively.
At the Fermi energy, the data is plotted for all times in Fig.~\ref{fig:fermi_inten_N2_N100_L10}.\label{fig:band_occupation_N2_N100_L10}}
\end{figure}

\begin{figure}
\centering
\includegraphics[width=0.8\textwidth]{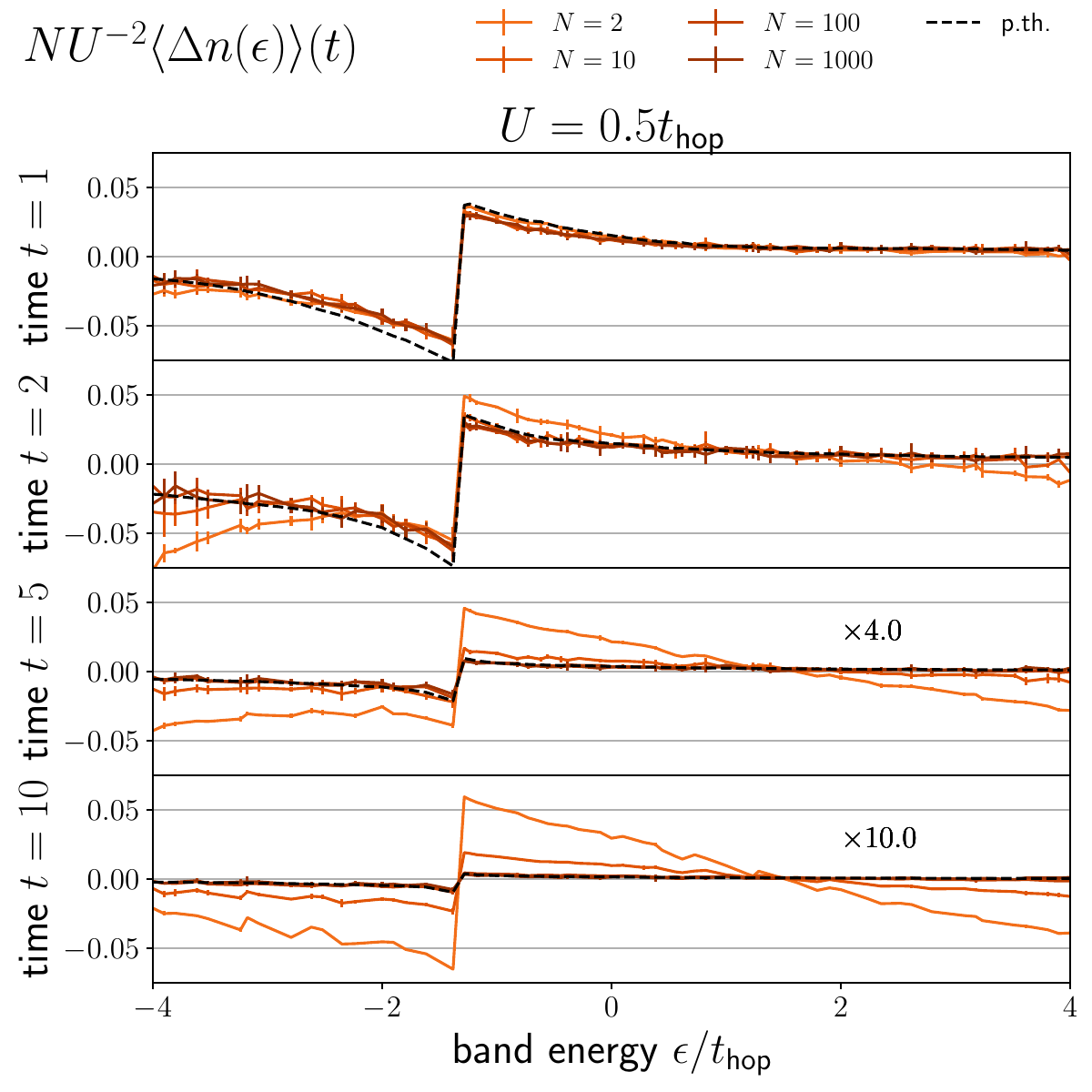}%
\caption{Electronic occupation numbers as a function of the band energy $\epsilon_k$ for a $20 \times 20$ square lattice Hubbard model at quarter filling ($N = 101$ particles) after a quench to interaction strengths $U = 0.5 t_\textrm{\scriptsize h}$.
Results for varying values of the fermion degeneracy $N$ are compared.
Error bars are obtained by calculating how well the numerical data restores spatial symmetries of the lattice.
At the Fermi energy, the data is plotted for all times in Fig.~\ref{fig:fermi_inten_U5e-1_L20}.\label{fig:band_occupation_U5e-1_L20}}
\end{figure}

The important next question is whether also the thermalization dynamics can be obtained within the semiclassical scheme.
At first glance, the departure from the plateau, e.g., for $N = 2$ in Fig.~\ref{fig:fermi_inten_U5e-1_L20}, seems to be consistent with the expected behavior.
However, turning to the interaction energy, 
we find that the departure goes hand in hand with a significant decrease of the interaction energy (for low values of $N$, $E_\textrm{\scriptsize int}$ even starts to decrease before the dynamics of $\Delta n_{k_F}$ away from the plateau becomes clearly visible).
Since local quantities like $E_\textrm{\scriptsize int}$ are expected to already relax to their thermal values in the prethermal regime,
the observed change of the interaction energy is unphysical behavior.
In contrast, for $N \geq 100$,
the interaction energy remains constant after prethermalization for all times considered here.
To shed more light on the dynamics beyond prethermalization, we consider the change of the full occupation number distribution $\Delta n(\epsilon, t) = n(\epsilon, t) - n(\epsilon, 0)$ for all single-particle energies $\epsilon_k$.
At time $t = 0$, $n(\epsilon, 0) = \theta(\epsilon_F - \epsilon)$.
Figure~\ref{fig:band_occupation_N2_N100_L10} shows $\Delta n(\epsilon, t)$ for the same set of parameters as in figure~\ref{fig:fermi_inten_N2_N100_L10} and figure~\ref{fig:band_occupation_U5e-1_L20} for the same parameters as in figure~\ref{fig:fermi_inten_U5e-1_L20}.
The error bars show the error estimate calculated with the procedure explained at the beginning of this section.
All data sets remain close to the perturbative result at time $t = 1$ (in units of $t_\textrm{\scriptsize h}^{-1}$).
However, especially for $N = 2$, one can find strong deviations from the perturbation theory and, most strikingly, negative occupation numbers (as well as occupation numbers larger than one) develop.
The semiclassical approximation is based on equations of motion for classical variables, which do not obey Pauli's exclusion principle.
Therefore, such unphysical behavior is indeed possible and clearly indicates the end of the range of validity of the method.
The results at late times (for instance, time $t = 10$ and $N = 2$ in Fig.~\ref{fig:band_occupation_N2_N100_L10}) suggest a kind of “straight-line” distribution as the fixed point of the dynamics.
We have seen the development of such a linear distribution in many simulations but more systematic studies and a better analytical understanding of the stationary distributions under the fTWA dynamics are required before general statements can be made.

\begin{figure}
\centering
\includegraphics[width=0.8\textwidth]{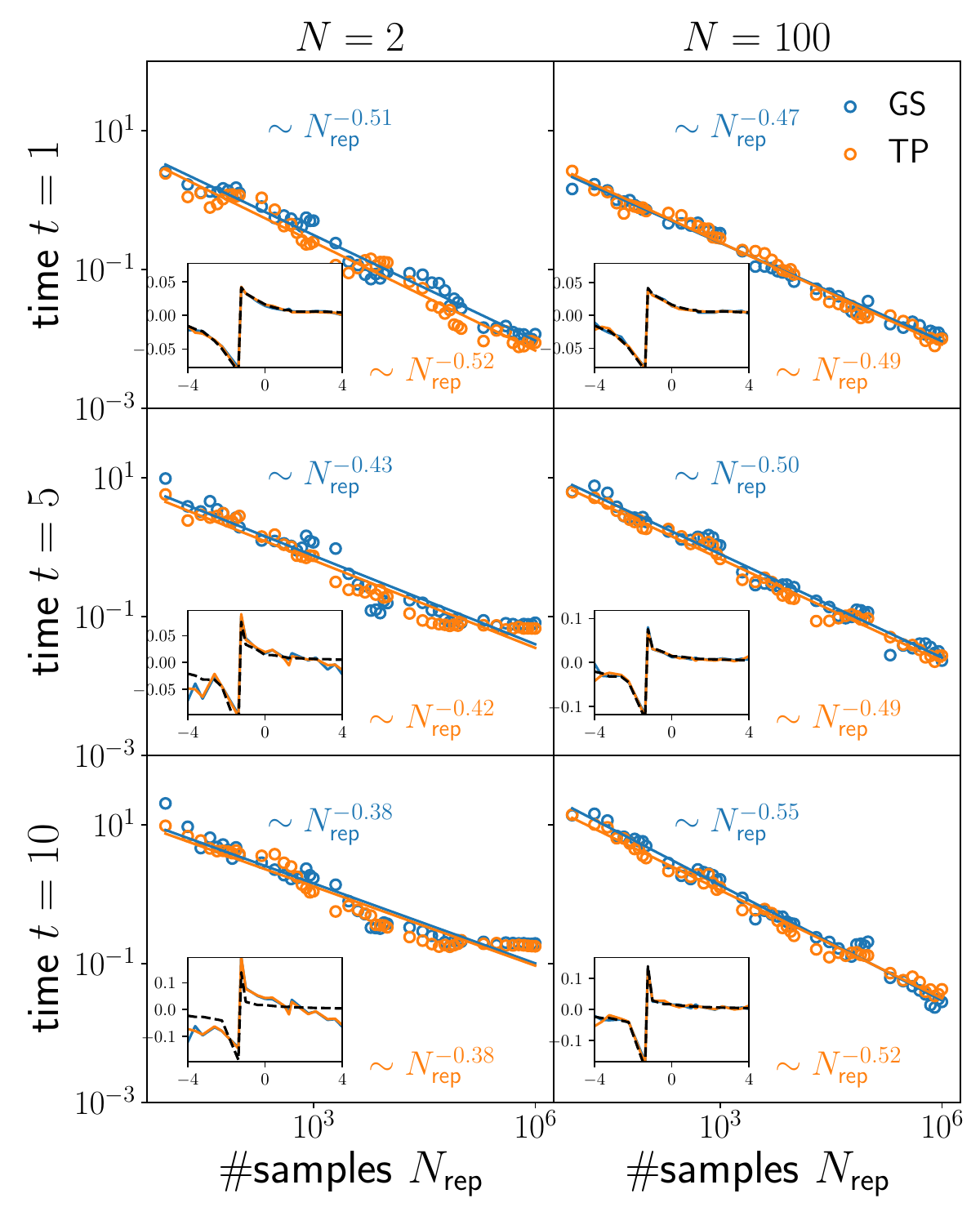}%
\caption{Deviation of the numerical fTWA data for $n(\epsilon_k)$ from the perturbative result for a quench to $U = 0.1 t_\textrm{\scriptsize h}$ in a $10 \times 10$ square lattice Hubbard model, calculated as the $L^2$-distance between both data sets.
The insets show the electronic occupation numbers at the respective instance of time $t$.
Two models for the initial state Wigner function are compared, a Gaussian distribution function (blue dots) and a two-point distribution function (orange dots), which yield very similar results.
In regimes, in which exact agreement with the perturbative result can be expected (early times and/or large $N$), the data shows a clear $\sim 1/\sqrt{N_\textrm{\scriptsize rep}}$ scaling of the error as a function of the number of trajectories.
\label{fig:rep_scaling_L10_U1e-1_T10}}
\end{figure}

Lastly, we would like to discuss figure~\ref{fig:rep_scaling_L10_U1e-1_T10}, which shows the deviation of numerical fTWA data for $U = 0.1 t_\textrm{\scriptsize h}$ from the perturbative result as a function of the number of trajectories $N_\textrm{\scriptsize rep}$.
In addition, we compare two Wigner function models, the Gaussian used so far and the two-point distribution function.
The deviation is calculated as the $L^2$-distance $\big( \int_{\epsilon_\textrm{\scriptsize min}}^{\epsilon_\textrm{\scriptsize max}} \textrm{d}\epsilon\; | n_1(\epsilon) - n_2(\epsilon) |^2 \big)^{1/2}$.
It is immediately clear from the data that for the interaction quench problem discussed in this paper both distribution functions yield identical results.
We find a scaling $\sim 1/\sqrt{N_\textrm{\scriptsize rep}}$ of the deviation in regimes in which the numerical data can be expected to be very close to the perturbative result, in particular for $N = 100$.
Such a scaling is in line with the law of large numbers because all samples are drawn independently at the initial time.

\section{Discussion and Conclusion}

In this paper we adopted the fermion degeneracy $N$ as a natural semiclassical expansion parameter, combined it with the fTWA method, and applied it to the well-understood problem of the interaction quench in the Hubbard model.
This allowed us to analyze the range of validity of fTWA in a systematic way.
%Our analytical and numerical results showed that fTWA correctly describes the electronic dephasing dynamics leading to prethermalization.
We conclude that in the regime of weak to moderate interaction strengths the method correctly reproduces the quantum dynamics at order $\sim \hbar = \frac{1}{N}$ and is valid at least up to and including the prethermalization regime.
The dynamics beyond prethermalization suffers from the development of negative occupation numbers and becomes unphysical.
A determination of the fixed point distributions under the fTWA dynamics will be a suitable starting point for further method development.
Nevertheless, fTWA is already a powerful tool for applications since it allows for a straightforward application to explicitly time-dependent problems~\cite{Alexander2022,Osterkorn2022,Paprotzki2023} or disordered models on large lattices~\cite{Sajna2020,Iwanek2023,Kaczmarek2023}.
Formulating SU($N$)-symmetric generalizations of lattice models, which are of interest for applications, allows one to choose the value of $N$ large enough so that the onset of the unphysical dynamics is pushed to irrelevantly late times.
In this way, corrections to the $\sim \hbar^0 = 1$ mean-field dynamics can be studied systematically.
A general advantage of fTWA is that the number of dynamical variables increases only quadratically with the system size, which allows the simulation of 2d lattice systems with a much larger number of sites possible than with exact diagonalization or tensor-network based approaches.
In addition, no memory kernels need to be tracked during the time evolution.
A downside, as mentioned in the text, is that the convergence with the number of trajectories can be slow at late times.
In any case, a thorough understanding of what the fTWA method can describe and what not is a necessary prerequisite for large-scale applications.

Finally, let us close with a few concrete ideas for future method development.
Refined Wigner functions like the two-point function (reminiscent of discrete TWA methods \cite{Schachenmayer2015}) can potentially increase the predictive power of fTWA~\cite{Ulgen2019}.
Although they did not yield any improvement for the interaction quench problem, other Wigner function models should be explored.
Another possible route for improvement is to add more complexity to the equations of motion~\cite{Czuba2020} by adding new variables.
Within boson and spin TWA, such an approach has already shown to yield improved results~\cite{Davidson2015,Wurtz2018}.

% To elucidate this issue it would be interesting to include further quantum corrections to TWA in the simulation \cite{Polkovnikov2010} or to use an optimized basis for the TWA setup.
% In addition, the assumption of a Gaussian Wigner function needs to be reviewed carefully, in particular with respect to the question up to which order the correlations of the initial product state are correctly described.

% It is hence a useful method if its range of validity is carefully analyzed.
% Starting from a mean-field initial state at $N \rightarrow \infty$, fTWA correctly describes the quantum dephasing that leads to the decay of coherent mean-field oscillations.
% The quantum scattering dynamics that ultimately results in thermalization is not contained in the present formulation of the method though.

\begin{ack}
We acknowledge helpful discussions with Anatoli Polkovnikov. This work was funded by the Deutsche Forschungsgemeinschaft (DFG, German Research Foundation) - 217133147/SFB 1073, project B07.
\end{ack}

\appendix

\section{Details on the perturbative calculation\label{app:perturb}}

Since $i \partial_t \tilde{\rho}_{kl} = \mathcal{O}(U)$, $i \partial_t \tilde{\rho}^{(0)}_{kl} = 0$ follows immediately.
Consequently, $\tilde{\rho}^{(0)}_{kl} = \delta_{k l} n_k(0)$.
The equation of motion for the $\mathcal{O}(U)$ contribution is
\begin{equation}
 i \partial_t \tilde{\rho}^{(1)}_{kl} = \frac{2}{V} \sum_{sp} \bigg[ \textrm{e}^{i \Delta\epsilon_{psl} t} \tilde{\rho}^{(0)}_{p+s-l,p} \tilde{\rho}^{(0)}_{ks}
 - \textrm{e}^{i \Delta\epsilon_{pks} t} \tilde{\rho}^{(0)}_{p+k-s,p} \tilde{\rho}^{(0)}_{sl} \bigg] .
 \label{eq:su_n_ftwa_eom_mom_int_1}
\end{equation}

It is possible to integrate the time-dependencies explicitly using the integral
\begin{equation}\begin{array}{*2{>{\displaystyle}l}}
&\mathcal{I}_1(\Delta\epsilon_{p a b}) = -i \int_0^t \textrm{d} t' \textrm{e}^{i \Delta\epsilon_{p a b} t'} \\
&\qquad = 
\left\{
\begin{array}{*2{>{\displaystyle}l}}
\frac{1}{\Delta\epsilon_{p a b}} \left( \textrm{e}^{i \Delta\epsilon_{p a b} t} - 1 \right), & \Delta\epsilon_{p a b} \neq 0.\\
-i t, & \Delta\epsilon_{p a b} = 0.
\end{array}
\right.
\end{array}\end{equation}
The Wigner function averages are performed manually using
\begin{equation}
 \langle \tilde\rho^{(0)}_1 \tilde\rho^{(0)}_2 \rangle = \langle \tilde\rho^{(0)}_1 \tilde\rho^{(0)}_2 \rangle^c + \langle \tilde\rho^{(0)}_1 \rangle \langle \tilde\rho^{(0)}_2 \rangle
\end{equation} and the initial data in \eqref{eq:ftwa_gauss_wigner_params}.
The structure of \eqref{eq:su_n_ftwa_eom_mom_int_1} is such that both terms cancel each other after the Wigner function averaging. Thus $\tilde{\rho}^{(1)}_{kl}(t) = 0$.

The next order $\mathcal{O}(U^2)$ already contains eight terms
\begin{equation}\begin{array}{*2{>{\displaystyle}l}}
&\tilde \rho^{(2)}_{k, l}(t) = \frac{4}{V^2} \sum_{sps'p'} \Big[ \\
&\mathcal{I}_2(\Delta\epsilon_{p's's}, \Delta\epsilon_{psl}) \tilde\rho^{(0)}_{(p+s-l), p} \tilde\rho^{(0)}_{(p'+s'-s), p'} \tilde\rho^{(0)}_{k, s'} \\
 &- \mathcal{I}_2(\Delta\epsilon_{p'ks'}, \Delta\epsilon_{psl}) \tilde\rho^{(0)}_{(p+s-l), p} \tilde\rho^{(0)}_{(p'+k-s'), p'} \tilde\rho^{(0)}_{s', s} \\
 &+ \mathcal{I}_2(\Delta\epsilon_{p's'p}, \Delta\epsilon_{psl}) \tilde\rho^{(0)}_{(p'+s'-p), p'} \tilde\rho^{(0)}_{(p+s-l), s'} \tilde\rho^{(0)}_{k, s} \\
 &- \mathcal{I}_2(\Delta\epsilon_{p'(p+s-l)s'}, \Delta\epsilon_{psl}) \tilde\rho^{(0)}_{(p'+p+s-l-s'), p'} \tilde\rho^{(0)}_{s', p} \tilde\rho^{(0)}_{k, s} \\
 &- \mathcal{I}_2(\Delta\epsilon_{p's'l}, \Delta\epsilon_{pks}) \tilde\rho^{(0)}_{(p+k-s), p} \tilde\rho^{(0)}_{(p'+s'-l), p'} \tilde\rho^{(0)}_{s, s'} \\
 &+ \mathcal{I}_2(\Delta\epsilon_{p'ss'}, \Delta\epsilon_{pks}) \tilde\rho^{(0)}_{(p+k-s), p} \tilde\rho^{(0)}_{(p'+s-s'), p'} \tilde\rho^{(0)}_{s', l} \\
 &- \mathcal{I}_2(\Delta\epsilon_{p's'p}, \Delta\epsilon_{pks}) \tilde\rho^{(0)}_{(p'+s'-p), p'} \tilde\rho^{(0)}_{(p+k-s), s'} \tilde\rho^{(0)}_{s, l} \\
 &+ \mathcal{I}_2(\Delta\epsilon_{p'(p+k-s)s'}, \Delta\epsilon_{pks}) \tilde\rho^{(0)}_{(p'+p+k-s-s'), p'} \tilde\rho^{(0)}_{s', p} \tilde\rho^{(0)}_{s, l} \Big]
 \label{eq:su_n_ftwa_eom_mom_int_2}
\end{array}\end{equation}
where
\begin{equation}
 \mathcal{I}_2(\Delta\epsilon_{p'ab}, \Delta\epsilon_{pcd}) = - \int_0^t \textrm{d} t' \int_0^{t'} \textrm{d} t'' \textrm{e}^{i \Delta\epsilon_{pcd} t'} \textrm{e}^{i \Delta\epsilon_{p'ab} t''} .
\end{equation}

The third moments of the Wigner function are evaluated using Wick's theorem for a Gaussian distribution \begin{equation}\begin{array}{*2{>{\displaystyle}l}}
 &\langle \tilde\rho^{(0)}_1 \tilde\rho^{(0)}_2 \tilde\rho^{(0)}_3 \rangle = \langle \tilde\rho^{(0)}_1 \rangle \langle \tilde\rho^{(0)}_2 \tilde\rho^{(0)}_3 \rangle^c + \langle \tilde\rho^{(0)}_2 \rangle \langle \tilde\rho^{(0)}_1 \tilde\rho^{(0)}_3 \rangle^c \\
 &\qquad + \langle \tilde\rho^{(0)}_3 \rangle \langle \tilde\rho^{(0)}_1 \tilde\rho^{(0)}_2 \rangle^c + \langle \tilde\rho^{(0)}_1 \rangle \langle \tilde\rho^{(0)}_2 \rangle \langle \tilde\rho^{(0)}_3 \rangle
 \label{eq:ftwa_gauss_wigner_third_moment}
\end{array}\end{equation}

It turns out that after averaging over the Wigner function \eqref{eq:su_n_ftwa_eom_mom_int_2} has the structure
\begin{equation}\begin{array}{*2{>{\displaystyle}l}}
 &\tilde\rho^{(2)}_{k l}(t) = \delta_{k, l} \frac{4}{N V^2} \sum_{pp'} J_{pp'k} \cdot \\
 &\quad \cdot \Big[ \mathcal{I}_2(- \Delta\epsilon_{pp'k}, \Delta\epsilon_{pp'k}) + \mathcal{I}_2(\Delta\epsilon_{pp'k}, -\Delta\epsilon_{pp'k}) \Big]
\end{array}\end{equation}
with
\begin{equation}\begin{array}{*2{>{\displaystyle}l}}
 &\mathcal{I}_2(\Delta\epsilon_{pp'k}, -\Delta\epsilon_{pp'k}) + \mathcal{I}_2(\leftrightarrow) = \\
 &\quad = \left\{
 \begin{array}{*2{>{\displaystyle}l}}
 -\frac{4}{(\Delta\epsilon_{pp'k})^2} \sin^2 \left( \frac{\Delta\epsilon_{pp'k}}{2} t \right) & \Delta\epsilon_{pp'k} \neq 0 \\ -t^2 & \Delta\epsilon_{pp'k} = 0
 \end{array}
 \right. .
\end{array}\end{equation}
These calculations finally yield \eqref{eq:ftwa_perturb_res}.
As discussed in the main text the elastic contributions that yield secular terms $\sim t^2$ need to be shifted to the unperturbed part of the Hamiltonian.

\section*{References}
\bibliography{refs}

\providecommand{\newblock}{}
\begin{thebibliography}{10}
\expandafter\ifx\csname url\endcsname\relax
  \def\url#1{{\tt #1}}\fi
\expandafter\ifx\csname urlprefix\endcsname\relax\def\urlprefix{URL }\fi
\providecommand{\eprint}[2][]{\url{#2}}

\bibitem{Polkovnikov2011}
Polkovnikov A, Sengupta K, Silva A and Vengalattore M 2011 {\em Rev. Mod.
  Phys.\/} {\bf 83} 863--883

\bibitem{Gross2017}
Gross C and Bloch I 2017 {\em Science\/} {\bf 357} 995--1001 ISSN 0036-8075,
  1095-9203

\bibitem{Dombi2020}
Dombi P, P{\'a}pa Z, Vogelsang J, Yalunin S~V, Sivis M, Herink G, Sch{\"a}fer
  S, Gro{\ss} P, Ropers C and Lienau C 2020 {\em Rev. Mod. Phys.\/} {\bf 92}
  025003

\bibitem{Manzeli2017}
Manzeli S, Ovchinnikov D, Pasquier D, Yazyev O~V and Kis A 2017 {\em Nature
  Reviews Materials\/} {\bf 2} 1--15 ISSN 2058-8437

\bibitem{Freericks2009a}
Freericks J~K, Krishnamurthy H~R and Pruschke T 2009 {\em Phys. Rev. Lett.\/}
  {\bf 102} 136401

\bibitem{Freericks2015}
Freericks J~K, Krishnamurthy H~R, Sentef M~A and Devereaux T~P 2015 {\em Phys.
  Scr.\/} {\bf T165} 014012 ISSN 1402-4896

\bibitem{Eckstein2009}
Eckstein M, Hackl A, Kehrein S, Kollar M, Moeckel M, Werner P and Wolf F 2009
  {\em Eur. Phys. J. Spec. Top.\/} {\bf 180} 217--235 ISSN 1951-6401

\bibitem{Paeckel2019}
Paeckel S, K{\"o}hler T, Swoboda A, Manmana S~R, Schollw{\"o}ck U and Hubig C
  2019 {\em Annals of Physics\/} {\bf 411} 167998 ISSN 0003-4916

\bibitem{Freericks2006}
Freericks J~K, Turkowski V~M and Zlati{\'c} V 2006 {\em Phys. Rev. Lett.\/}
  {\bf 97} 266408

\bibitem{Aoki2014}
Aoki H, Tsuji N, Eckstein M, Kollar M, Oka T and Werner P 2014 {\em Rev. Mod.
  Phys.\/} {\bf 86} 779--837

\bibitem{Hackl2008}
Hackl A and Kehrein S 2008 {\em Phys. Rev. B\/} {\bf 78} 092303

\bibitem{Sinatra2002}
Sinatra A, Lobo C and Castin Y 2002 {\em J. Phys. B: At. Mol. Opt. Phys.\/}
  {\bf 35} 3599--3631 ISSN 0953-4075

\bibitem{Polkovnikov2002}
Polkovnikov A, Sachdev S and Girvin S~M 2002 {\em Phys. Rev. A\/} {\bf 66}
  053607

\bibitem{Polkovnikov2003}
Polkovnikov A 2003 {\em Phys. Rev. A\/} {\bf 68} 053604

\bibitem{Ayik2008}
Ayik S 2008 {\em Physics Letters B\/} {\bf 658} 174--179

\bibitem{Lacroix2014}
Lacroix D, Hermanns S, Hinz C~M and Bonitz M 2014 {\em Phys. Rev. B\/} {\bf 90}
  125112

\bibitem{Lacroix2014a}
Lacroix D and Ayik S 2014 {\em The European Physical Journal A\/} {\bf 50}
  1--34

\bibitem{Davidson2017}
Davidson S~M, Sels D and Polkovnikov A 2017 {\em Annals of Physics\/} {\bf 384}
  128--141 ISSN 0003-4916

\bibitem{Moeckel2008}
Moeckel M and Kehrein S 2008 {\em Phys. Rev. Lett.\/} {\bf 100} 175702

\bibitem{Moeckel2009}
Moeckel M and Kehrein S 2009 {\em Annals of Physics\/} {\bf 324} 2146--2178
  ISSN 0003-4916

\bibitem{Moeckel2010}
Moeckel M and Kehrein S 2010 {\em New J. Phys.\/} {\bf 12} 055016 ISSN
  1367-2630

\bibitem{Hamerla2014}
Hamerla S~A and Uhrig G~S 2014 {\em Phys. Rev. B\/} {\bf 89} 104301

\bibitem{Yaffe1982}
Yaffe L~G 1982 {\em Rev. Mod. Phys.\/} {\bf 54} 407--435

\bibitem{Bickers1987}
Bickers N~E 1987 {\em Rev. Mod. Phys.\/} {\bf 59} 845--939

\bibitem{Gardiner1997}
Gardiner C~W 1997 {\em Phys. Rev. A\/} {\bf 56} 1414--1423

\bibitem{Polkovnikov2010}
Polkovnikov A 2010 {\em Annals of Physics\/} {\bf 325} 1790--1852 ISSN
  0003-4916

\bibitem{Schmitt2019}
Schmitt M, Sels D, Kehrein S and Polkovnikov A 2019 {\em Phys. Rev. B\/} {\bf
  99} 134301

\bibitem{Sajna2020}
Sajna A~S and Polkovnikov A 2020 {\em Phys. Rev. A\/} {\bf 102} 033338

\bibitem{Iwanek2023}
Iwanek {\L}, Mierzejewski M, Polkovnikov A, Sels D and Sajna A~S 2023 {\em
  Physical Review B\/} {\bf 107} 064202

\bibitem{Kaczmarek2023}
Kaczmarek A and Sajna A~S 2023 {\em Physical Review B\/} {\bf 108} 134304

\bibitem{Ulgen2019}
Ulgen I, Yilmaz B and Lacroix D 2019 {\em Phys. Rev. C\/} {\bf 100} 054603

\bibitem{Read1991}
Read N and Sachdev S 1991 {\em Phys. Rev. Lett.\/} {\bf 66} 1773--1776

\bibitem{Sachdev1991}
Sachdev S and Read N 1991 {\em Int. J. Mod. Phys. B\/} {\bf 05} 219--249 ISSN
  0217-9792

\bibitem{Newns1987}
Newns D~M and Read N 1987 {\em Advances in Physics\/} {\bf 36} 799--849 ISSN
  0001-8732

\bibitem{Affleck1988}
Affleck I and Marston J~B 1988 {\em Phys. Rev. B\/} {\bf 37} 3774--3777

\bibitem{Osterkorn2022}
Osterkorn A and Kehrein S 2022 {\em Physical Review B\/} {\bf 106} 214318

\bibitem{Kronenwett2011}
Kronenwett M and Gasenzer T 2011 {\em Appl. Phys. B\/} {\bf 102} 469--488 ISSN
  1432-0649

\bibitem{Weidinger2017}
Weidinger S~A and Knap M 2017 {\em Sci Rep\/} {\bf 7} 45382 ISSN 2045-2322

\bibitem{Walz2018}
Walz R, Boguslavski K and Berges J 2018 {\em Phys. Rev. D\/} {\bf 97} 116011

\bibitem{Gorshkov2010}
Gorshkov A~V, Hermele M, Gurarie V, Xu C, Julienne P~S, Ye J, Zoller P, Demler
  E, Lukin M~D and Rey A~M 2010 {\em Nature Phys\/} {\bf 6} 289--295 ISSN
  1745-2481

\bibitem{Choudhury2020}
Choudhury S, Islam K~R, Hou Y, Aman J~A, Killian T~C and Hazzard K~R~A 2020
  {\em Phys. Rev. A\/} {\bf 101} 053612

\bibitem{Marston1989}
Marston J~B and Affleck I 1989 {\em Phys. Rev. B\/} {\bf 39} 11538--11558

\bibitem{Ahnert2011}
Ahnert K and Mulansky M 2011 {\em AIP Conference Proceedings\/} {\bf 1389}
  1586--1589 ISSN 0094-243X

\bibitem{Sanderson2016}
Sanderson C and Curtin R 2016 {\em The Journal of Open Source Software\/} {\bf
  1} 26

\bibitem{Sanderson2018}
Sanderson C and Curtin R~R 2018 A {{User-Friendly Hybrid Sparse Matrix Class}}
  in {{C}}++ {\em {{ICMS}}\/}

\bibitem{Schubert2018}
Schubert E and Gertz M 2018 Numerically stable parallel computation of
  (co-)variance {\em Proceedings of the 30th {{International Conference}} on
  {{Scientific}} and {{Statistical Database Management}}\/} {{SSDBM}} '18
  ({Bozen-Bolzano, Italy}: {Association for Computing Machinery}) pp 1--12 ISBN
  978-1-4503-6505-5

\bibitem{Alexander2022}
Alexander M and Kollar M 2022 {\em physica status solidi (b)\/} {\bf 259}
  2100280

\bibitem{Paprotzki2023}
Paprotzki E, Osterkorn A, Misha V and Kehrein S 2023 {\em arXiv preprint
  arXiv:2312.12291\/}

\bibitem{Schachenmayer2015}
Schachenmayer J, Pikovski A and Rey A~M 2015 {\em Phys. Rev. X\/} {\bf 5}
  011022

\bibitem{Czuba2020}
Czuba T, Lacroix D, Regnier D, Ulgen I and Yilmaz B 2020 {\em Eur. Phys. J.
  A\/} {\bf 56} 111 ISSN 1434-601X

\bibitem{Davidson2015}
Davidson S~M and Polkovnikov A 2015 {\em Phys. Rev. Lett.\/} {\bf 114} 045701

\bibitem{Wurtz2018}
Wurtz J, Polkovnikov A and Sels D 2018 {\em Annals of Physics\/} {\bf 395}
  341--365 ISSN 00034916 (\textit{Preprint} \eprint{1804.10217})

\end{thebibliography}

\end{document}